\documentclass{aa}
\usepackage{graphicx,natbib}
\bibpunct{(}{)}{;}{a}{}{,}

\def\aj{AJ}
\def\apj{ApJ}

\def\aap{A\&A}

\begin{document}

\sloppypar

%
\title{Why the braking indices of young pulsars are less than 3?}

\author{W. C. Chen and X. D. Li }

\institute{Department of Astronomy, Nanjing University, Nanjing 210093, China\\
\email{chenwc@nju.edu.cn,lixd@nju.edu.cn} }
\date{}
\authorrunning{Chen \& Li }

\abstract{In this letter we discuss two possible reasons which
cause the observed braking indices n of young radio pulsars to be
smaller than 3: (a) the evolving spin-down model of the magnetic
field component $B_{\perp}$ increases with time; (b) the extrinsic
braking torque model in which the tidal torques exerted on the
pulsar by the fallback disk, and carries away the spin angular
momentum from the pulsar. Based on some simple assumptions, we
derive the expression of the braking indices, and calculate the
spin-down evolutionary tracks of pulsars for different input
parameters.
   }

   \maketitle

%

\section{Introduction}
Since the first radio pulsar was discovered by \citet{hew68},
continued radio observations at several wavelengths have not only
increased the number of detected pulsars but also amassed a wealth
of details about the temporal, spectral and polarization
properties of the observed pulses. It was then quickly established
that radio pulsars are rapidly rotating strongly magnetized
neutron stars, powered by their rotational kinetic energy and
losing energy by magnetic dipole radiation \citep{Gol68}. The spin
angular velocity of pulsars decreases with time, and the spin-down
may be written by the relation \citep{man77},
\begin{equation}
I\dot{\Omega}=-K\Omega^{n},
\end{equation}
where $\Omega$ is the spin angular velocity, K is a positive
parameter depending on the magnetic dipole moment, and n is the
braking index. The braking index n can be a measured for young
radio pulsars by differentiating equation (1),
\begin{equation}
n=\frac{\ddot{\Omega}\Omega}{\dot{\Omega}^{2}}.
\end{equation}

If the spin-down torque entirely results from magnetic dipole
radiation with a constant magnetic field, the predicted braking
index is $\rm n=3$. However, if the braking is due to a
relativistic stellar wind, the braking index $\rm n=1$
\citep{mic69}. The observed braking index n, which has been
determined for several young radio pulsars, are all less than 3.
For example, the braking index of the Crab pulsar PSR 0531+21 is
2.509 $\pm$ 0.001 \citep{lyn93}, n=2.140 $\pm$ 0.009 for PSR
0540-69 \citep{liv05}, n=2.837 $\pm$ 0.001 for PSR 1509-58
\citep{kas94}, n=2.91 $\pm$ 0.05 for PSR 1119-6127, $\rm n=1.4 \pm
0.2$ for the Vela pulsar PSR 0833-45 \citep{lyn96}, and $\rm
n=2.65 \pm 0.01$ for PSR J1846-0258 which located at the center of
supernova remnant Kes 75 \citep{liv06}.

In previous works, many authors proposed various models to explain
the discrepancy between observed and theoretical predicted value
of n. \citet{pen82} suggested that neutrino and photon radiation
coming from superfluid neutrons may brake the pulsars.
\citet{bla88} proposed that $\rm n <3$ is due to a multipole field
and field evolution. \citet{men01} and \citet{alp02} presented a
model that both magnetic dipole radiation and the propeller torque
applied by the debris disk formed after supernova explosion can
cause spin-down of pulsars. \citet{xu01} and \citet{wu03} pointed
out that the combination of dipole radiation and the unipolar
generator may cause n $<$ 3. \citet{all97} suggested that a
variation of the torque function is important attribution for low
braking index, and so on.

In this letter, we suggest that the braking index may be less than
3 due to increasing of magnetic field strength of pulsars or
braking of fallback disks surrounding pulsars. In next section, we
describe two simple models to explain why the observed braking
index $\rm n<3$. In section 3 we present the calculated results
for the braking model of the fallback disk. In section 4 we make a
brief discussion and summary.

\section{Model}

Pulsar braking process is still poorly known even more than
several decades after it was discovered. The best radiation model
is the magnetic dipole radiation, one which has been proven to be
extremely successful in explaining the observed properties of
pulsars \citep{pac67,gun69}. The rotating pulsars are braked by
magnetic dipole radiation \citep{ost69} or particle outflow from
the poles of a dipole field \citep{Gol69} which has a constant
field strength. However, many works propose that there are other
energy loss mechanisms and braking torques besides magnetic dipole
radiation,
\begin{equation}
I\dot{\Omega} =-K\Omega^{3}+T(\Omega),
\end{equation}
where $K=2B^{2}R^{6}sin^{2}\theta /3c^{3}$; $\theta$ is the
inclination of the magnetic axis with respect to the rotation
axis; $I,B, R$ are the momentum of inertia, the surface magnetic
field strength, and the radius of the pulsar, respectively; c is
the velocity of light, and $T(\Omega)$ is other braking torque.

We suggest that the observed braking index is less than 3 possibly
due to the following two reasons: (a) $K=K(\Omega)$,
$T(\Omega)=0$, i.e. the other braking torque does not exist except
for magnetic dipole radiation; (b) $K=\rm constant$,
$T(\Omega)\neq 0$.

\subsection{Case A: $K=K(\Omega), T(\Omega)=0$}

If the momentum of inertia of the pulsar is a constant, by
differentiating equation (3), from equation (2) we can obtain the
observed braking index
\begin{equation}
n=3+\frac{d\rm ln K}{d\rm ln\Omega}.
\end{equation}
If the second term on the right-hand side of the above equation is
negative, therefore $n<3$ for young pulsars.

Assuming that the surface dipole magnetic field changes with
angular velocity due to some unknown physical mechanisms, $R$, and
the braking index n is constant during the pulsar spin-down, we
can derive the evolution of magnetic field component $B_{\perp}=B
\rm sin \theta$ in the direction perpendicular to the rotation
axis with spin angular velocity of pulsar from equation (4)
\begin{equation}
B_{\perp} =C\Omega^{\frac{n-3}{2}},
\end{equation}
where $C$ is the integrating constant determined by the parameters
of the pulsar. Setting the initial spin angular velocity is
$\Omega_{0}$, and the magnetic field component is $B_{\perp,0}$
when pulsar was born, then
\begin{equation}
B_{\perp}=B_{\perp,0}\left(\frac{\Omega}{\Omega_{0}}\right)^{\frac{n-3}{2}}.
\end{equation}

From equation (6), the magnetic field component $B_{\perp}$ is
constant during spin down if $\rm n=3$. For young radio pulsars,
$n<3$, $\Omega<\Omega_{0}$ due to spin-down, we therefore can
obtain $B_{\perp}>B_{\perp,0}$. Namely, the magnetic field
vertical component $B_{\perp}$ of young pulsars would increase if
the braking torque is only magnetic dipole radiation and the
assumption mentioned in the above is true \citep{bla88}.

\subsection{Case B: $K=\rm constant, T(\Omega)\neq0$}
Actually, other braking torque possibly exist besides magnetic
dipole radiation. \citet{mic88} proposed that there may be disks
surrounding radio pulsars, produced by supernova fallback.
\citet{men01} applied the fallback disk model and the propeller
torque to explain the lower braking index of young radio pulsars.
Following this idea we assume that there is a fallback disk
residing outside the light cylinder of young pulsars, and tidal
torques instead traditional propeller torques are put on the
fallback disk, and carry away angular momentum from the pulsar
inside it.

Same as standard thin disk, the viscous torque in the fallback
disk is \citep{Shak73}
\begin{equation}
T=2\pi r\nu\Sigma r\frac{d\Omega_{r}}{dr}r,
\end{equation}
where $\nu$ is the viscosity and $\Sigma$ is the surface density,
$\Omega_{r}=(GM/r^{3})^{1/2}$ is the local rotational angular
velocity in the fallback disk, and $M$ is the mass of pulsar.

If the gravitational interaction of the pulsar with the fallback
disk only occurs at the inner edge $r_{\rm i}$ of the fallback
disk, angular momentum from pulsar feed into the fallback disk at
a rate proportional to the surface density $\Sigma_{\rm i}$ at the
inner region of the fallback disk. As a result of continuous input
of mass at the inner edge of the fallback disk, the viscous torque
exerted on the pulsar by the fallback disk  can be shown to be
\citep{taam01,spru01}
\begin{equation}^{}
\dot{J}_{\rm d}=-\left(\frac{r_{\rm i}}{R}\right)^{1/2}\Omega
R^{2}\dot{M}\left(\frac{t}{t_{\rm vi}}\right)^{1/2},
\end{equation}
where, $t$ denotes the spin-down age of the pulsar, $\dot{M}$ is
the mass inflow rate (assumed to be constant here), $t_{\rm vi}$
is the viscous timescale at the inner edge of disk.

 Assuming that the fallback disk is hydrostatically
supported and geometrically thin with a pressure scale height to
radius ratio of $\beta=H_{\rm i}/r_{\rm i}$, we estimate the
viscosity $\nu_{\rm i}$ \citep{chen06} at the inner edge of the
fallback disk using standard $\alpha$ prescription presented by
\citet{Shak73}
\begin{equation}
\nu_{\rm i}=\alpha_{\rm SS}\beta^{2} \Omega R^{3/2}r_{\rm
i}^{1/2},
\end{equation}
where $\alpha_{\rm SS}$ is the viscosity parameter. In the
following calculations we set $\alpha_{\rm SS}=0.001$.

By simple algebra, from the above equations we obtain the viscous
timescale at the inner edge in the fallback disk,
\begin{equation}
t_{\rm vi}=\frac{4}{3\alpha_{\rm SS} \beta^{2}\Omega
}\left(\frac{r_{\rm i}}{R}\right)^{3/2}.
\end{equation}

The total braking torque consisting of magnetic dipole radiation
and the fallback disk braking can be written as
\begin{equation}
I\dot{\Omega}=-K\Omega^{3}+T_{\rm d}(\Omega),
\end{equation}
with
\begin{equation}
  T_{\rm d}(\Omega)=-\gamma^{-1}c^{-1/4}R^{9/4}\beta(3\alpha_{\rm
SS}t/4)^{1/2}\dot{M}\Omega^{7/4},
\end{equation}
where $\gamma=(r_{\rm i}/R_{\rm lc})^{1/4}$, $R_{\rm lc}=c/\Omega$
is the light-cylinder radius of the pulsar. We assume the inner
edge in fallback disk to be located at the light-cylinder radius
\citep{men01}, i.e. $\gamma = 1$.

Similar as in the above subsection, differentiating equation (11),
from equation (2) we can get
\begin{equation}
n=3-(\frac{5}{4}+\frac{\tau}{t})\eta,
\end{equation}
where $\tau=-\Omega/(2\dot{\Omega})$ is the canonical spin-down
age of pulsars, $\eta=T_{\rm d}(\Omega)/(-K\Omega^{3}+T_{\rm
d}(\Omega))$ is the efficiency of the disk braking torque. The
second term on the right-hand side of the above equation is
positive, resulting in the braking index less than 3.

\section{Calculated Results for Case B}

We have adopted the semi-analytical method to calculate the
evolution of $P,\dot{P}$, and n for radio pulsars with a fallback
disk. We set some typical values for pulsars,
$B_{\perp}=10^{12}G$, $M=1.4M_{\odot}$, $R=10^{6}\rm cm$, i.e.
$K\simeq 2.5\times10^{28}\rm  g\cdot cm^{2}s$, $I\approx
10^{45}\rm g\cdot cm^{2}$; and $\beta=0.02,0.03,0.04$
\citep{bell04} in the calculations. Setting the initial spin
period of the pulsar $P_{\rm i}=0.02\rm s$, we obtain the pulsar
evolution in the $\dot{P}-P$ diagram and the evolutionary tracks
of the braking index in Figs. 1 and 2 for different values of
$\dot{M}$ and $\beta$, respectively. We can find that the braking
index is all the way less than 3, and first increases then
decreases during the spin-down from Fig. 2. The larger $\dot{M}$
and $\beta$, the larger $\dot{P}$ and the smaller
 n.

 \begin{figure}
\centering
\includegraphics[angle=0,width=8cm]{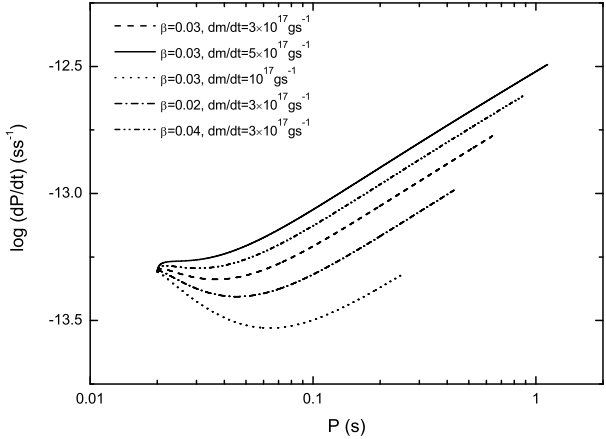}
\caption{The pulsar evolution in the $\dot{P}$ - $P$ diagram for
different input parameters $\beta$ and the constant mass inflow
rate $\dot{M}$ in the fallback disk. } \label{FigVibStab}
\end{figure}

\begin{figure}
\centering
\includegraphics[angle=0,width=7.5cm]{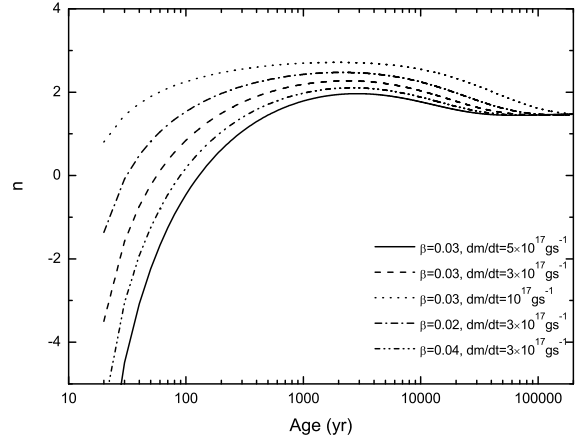}
\caption{Evolutionary tracks of braking index for different input
parameters $\beta$ and the constant mass inflow rate $\dot{M}$ in
the fallback disk. } \label{FigVibStab}
\end{figure}

 For the Crab pulsar, we can derive the current braking
index since its age is a certain value in the case of a steady
inflow in fallback disk . Set $t=930 \rm yr$, the mass inflow rate
 $\dot{M}=3\times10^{17}\rm g\cdot s^{-1}$, and $\beta=0.03, 0.04, 0.05$, we derive the
braking index are $n=2.60,2.49,2.39$ from equation (13), which can
account for the measured value within $5\%$.

\section{Discussion and Summary}
If the radiation process of pulsars is pure magnetic dipole
radiation, and $I,M,R$ are constant, the change of magnetic field
strength is the possible reason of lower braking index. From
$\dot{\Omega}\propto \Omega^{\rm n}$, we can derive the
evolutionary tracks of four young pulsars in $\dot{P}-P$ diagram
(see Fig. 3). Anomalous X-ray pulsars (AXPs) and soft Gamma-ray
repeaters (SGRs) are a small class of pulsars with the relatively
narrow period range of $5 - 12 \rm s$ and a higher spin-down rates
\citep{man01}. The spin-down luminosity
$\dot{E}=4\pi^{2}I\dot{P}/P^{3}$ of AXPs and SGRs is much less
than their X-ray luminosity. \citet{tho96} suggested that they are
powered by decay of strong magnetic fields. It is clear that the
magnetic field strength of Vela pulsar can exceed $10^{14}$ G when
$P\sim 5-6 \rm s$ from Fig. 3, suggesting that magnetars as AXPs
and SGRs possibly evolve from normal radio pulsars as Vela
\citep{lin04}.

\begin{figure}
\centering
\includegraphics[angle=0,width=8cm]{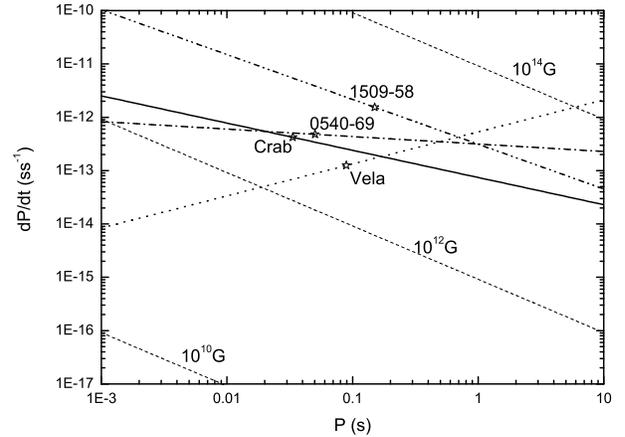}
\caption{Evolutionary tracks for four young radio pulsars, the
solid, dot, dashed-dot, and dashed-dot dot lines correspond to the
Crab pulsar, the Vela pulsar, PSR 0540-69, and PSR 1509-58,
respectively. The four open stars denote the present location of
four young pulsars. } \label{FigVibStab}
\end{figure}

However, the neutron star may lose spin angular momentum not only
by magnetic dipole radiation but also by other braking mechanisms.
Many authors proposed that young radio pulsars could be surrounded
by a remnant disk generated by supernova fallback
\citep{mic81,mic83}, which may contribute the braking torque
causing spin-down of neutron stars. Assuming that a constant rate
of material feeding into fallback disk, we derive the tidal torque
exerted on the pulsar by the fallback disk. Our calculations show
that the braking torque is significantly influenced by the input
parameters $\dot{M}$ and $\beta$. The larger values of $\dot{M}$
or $\beta$ can yield a larger braking torque and a smaller braking
index n. Fig. 1 shows that large $\dot{M}$ or $\beta$ can cause
the generation of pulsars with high spin-down rate, like SGRs and
AXPs. The evolution of the braking index shows that they are all
less than 3 during spin-down (see Fig. 2).

If some of the so-called magnetar, i.e. SGRs and AXPs, evolve from
normal young pulsars as described by case a, their surface
magnetic field should be strong as $\ga 10^{14}$ G, while in case
b they may have usual magnetic field $\sim 10^{12}$G but with high
spin-down rates. Measurement of cyclotron lines in magnetars may
present constrains on the two kinds of models. In addition, the
presence of fallback disks surrounding pulsars could be detected,
since a relatively dense fallback disk can produce optical and
infrared radiation. Optical emission is a common character of
young pulsars. The Crab pulsar, the Vela pulsar, and PSR 0540-69,
which possess low braking index, have been shown to possess a
pulsating optical counterpart \citep{men01}, implying the possible
existence of the fallback disks. Searching for optical and
infrared emission from the fallback disks surrounding young
pulsars will be of great importance for their evolution.

\begin{acknowledgements}
We are grateful for an anonymous referee for helpful comments on
this manuscript. This work was supported by the Natural Science
Foundation of China (NSFC) under grant number 10573010.
\end{acknowledgements}


\begin{thebibliography}{}


\bibitem[\protect\citeauthoryear{Allen \& Horvath} {1997}]{all97} Allen, M. P., \& Horvath, J. E.  1997, \apj, 488, 409
\bibitem[\protect\citeauthoryear{Alpar et al.} {2001}]{alp02} Alpar, M. A., Ankay, A.,\& Yazgan, E.  2001, \apj, 557, L61
\bibitem[\protect\citeauthoryear{Belle et al.} {2004}]{bell04} Belle, K. E., Sanghi, N., Howell, S. B., Holberg, J. B., \& Williams, P. T. 2004, \aj, 128, 448
\bibitem[\protect\citeauthoryear{Blandford \& Romani} {1988}]{bla88} Blandford, R. D., \& Romani, R. W. 1988, MNRAS, 234, 57
\bibitem[\protect\citeauthoryear{Chen et al.} {2006}]{chen06} Chen, W. C., Li, X. D., \& Wang, Z. R. 2006, PASJ, 58, 153
\bibitem[\protect\citeauthoryear{Gold} {1968}]{Gol68} Gold, T. 1968, Nature, 218, 731
\bibitem[\protect\citeauthoryear{Guliahorn \& Rankin} {1978}]{Gul78} Guliahorn, G. E., \& Rankin, J, M. 1978, \apj, 83, 10
\bibitem[\protect\citeauthoryear{Gunn \& Ostriker} {1969}]{gun69} Gunn, J. E., \& Ostriker, J. P. 1969, Nature, 221, 454
\bibitem[\protect\citeauthoryear{Goldreich \& Jullian} {1969}]{Gol69} Goldreich, P., \& Jullian, W, H. 1969, \apj, 157, 869
\bibitem[\protect\citeauthoryear{Hewish et al.} {1968}]{hew68} Hewish, A., Bell, S. J., Pilkington, J. D., Scott, P. F., \& Collins, R. A. 1968, Nature, 217, 709
\bibitem[\protect\citeauthoryear{Kaspi et al.} {1994}]{kas94} Kaspi, V. M., Manchester, R. N., Siegman, B., Johnston, S., \& Lyne, A. G.  1994, \apj, 422, L83
\bibitem[\protect\citeauthoryear{Lin \& Zhang } {2004}]{lin04} Lin, J. R., \& Zhang, S. N. 2004, \apj, 615, L133
\bibitem[\protect\citeauthoryear{Livingstone et al.} {2005}]{liv05} Livingstone, M. A., Kaspi, V. M., \& Gavriil, F. P. 2005, \apj, 633, 1095
\bibitem[\protect\citeauthoryear{Livingstone et al.} {2006}]{liv06} Livingstone, M. A., Kaspi, V. M., \& Gotthelf, E. V. 2006, submitted to \apj (astro-ph/0601530)
\bibitem[\protect\citeauthoryear{Lyne et al.} {1993}]{lyn93} Lyne, A. G., Pritchard, R. S.,\& Smith, F. G. 1993, MNRAS, 265, 1003
\bibitem[\protect\citeauthoryear{Lyne et al.} {1996}]{lyn96} Lyne, A. G., Pritchard, R. S., Graham-Smith, F., \& Camilo, F. 1996, Nature, 381, 497
\bibitem[\protect\citeauthoryear{Manchester } {2001}]{man01} Manchester, R. N. 2001, in AIP Conf. Proc. 565, 305
\bibitem[\protect\citeauthoryear{Manchester \& Peterson} {1989}]{man89} Manchester, R. N., \&  Peterson, B. A.  1989, \apj, 342, L23
\bibitem[\protect\citeauthoryear{Manchester \& Taylor} {1977}]{man77} Manchester, R. N., \& Taylor, J. H.  1977, pulsars, San Francisco, Freeman
\bibitem[\protect\citeauthoryear{Melatos} {1997}]{mel84} Melatos, A. 1997, MNRAS, 288, 1049
\bibitem[\protect\citeauthoryear{Menou et al.} {2001}]{men01} Menou, K., Perna, R., \& Hernquist, L.  2001, \apj, 554, L63
\bibitem[\protect\citeauthoryear{Michel} {1969}]{mic69} Michel, F. C. 1969, \apj, 158, 727
\bibitem[\protect\citeauthoryear{Michel \& Dessler} {1981}]{mic81} Michel, F. C., \& Dessler, A. J.  1981, \apj, 251, 654
\bibitem[\protect\citeauthoryear{Michel \& Dessler} {1983}]{mic83} Michel, F. C., \& Dessler, A. J.  1983, Nature, 303, 48
\bibitem[\protect\citeauthoryear{Michel} {1988}]{mic88} Michel, F. C. 1988, Nature, 333, 644
\bibitem[\protect\citeauthoryear{Ostriker \& Gunn} {1969}]{ost69} Ostriker, J. P., \& Gunn, J. E. 1969, \apj, 157, 1395
\bibitem[\protect\citeauthoryear{Pacini} {1967}]{pac67} Pacini, F. 1967, Nature, 216, 567
\bibitem[\protect\citeauthoryear{Peng et al.} {1982}]{pen82} Peng, Q. H., Huang, K. L., \& Huang, J. H. 1982, A\&A, 107, 258
\bibitem[\protect\citeauthoryear{Shakura \& Sunyaev}{1973}]{Shak73} Shakura, N. I., \& Sunyaev, R. A. 1973, \aap, 24, 337
\bibitem[\protect\citeauthoryear{Simon \& David} {1999}]{sim99} Simon, J., \& David, G.  1999, MNRAS, 306, L50
\bibitem[\protect\citeauthoryear{Spruit \& Tamm} {2001}]{spru01} Spruit, H. C., \& Taam, R. E. 2001, \apj, 548, 900
\bibitem[\protect\citeauthoryear{Taam \& Spruit} {2001}]{taam01} Taam, R. E., \& Spruit, H. C.  2001, \apj, 561, 329
\bibitem[\protect\citeauthoryear{Thompson \& Duncan} {1996}]{tho96} Thompson, C., \& Duncan, R. C. 1996, \apj, 473, 322
\bibitem[\protect\citeauthoryear{Wu et al.} {2003}]{wu03} Wu, F., Xu, R. X., \& Gil, J. 2003, \aap, 409, 641
\bibitem[\protect\citeauthoryear{Xu \& Qiao} {2001}]{xu01} Xu, R. X., \& Qiao, G. J. 2001, \apj, 561, L85

\end{thebibliography}
\end{document}